\documentclass{article}

\usepackage{arxiv}

\usepackage{amsmath}
\usepackage[nopatch]{microtype}
\usepackage{booktabs}
\usepackage{amsmath}
\usepackage{amsfonts}
\usepackage{amssymb}
\usepackage{booktabs}
\usepackage{graphicx}
\usepackage{subfig}
\usepackage{comment}
\usepackage{rotating} 
\usepackage{pdflscape}
\usepackage{xurl}
\usepackage[hidelinks]{hyperref}
\usepackage{amsfonts}
\usepackage[style=authoryear]{biblatex}
\usepackage[linesnumbered,ruled,vlined]{algorithm2e}

\DeclareFieldFormat{url}{}
\DeclareFieldFormat{urldate}{}
\DeclareFieldFormat{note}{}
\DeclareFieldFormat{doi}{}

\title{Spatial Modeling and Risk Zoning of Global Extreme Precipitation via Graph Neural Networks and $r$-Pareto Processes}

\author{Zimu Wang  \\
	College of Finance and Statistics\\
	Hunan University\\
	Changsha, Hunan Province, China, 410082 \\
	\texttt{wzm040218@hnu.edu.cn} \\
	\And
	Yifan Wu \\
	College of Finance and Statistics\\
	Hunan University\\
	Changsha, Hunan Province, China, 410082 \\
	\texttt{wuyifan04@hnu.edu.cn} \\
	\AND
	Daning Bi \thanks{Corresponding author} \\
	College of Finance and Statistics\\
	Hunan University\\
	Changsha, Hunan Province, China, 410082 \\
	\texttt{daningbi@hnu.edu.cn} \\
}

\date{}

\renewcommand{\headeright}{}
\renewcommand{\undertitle}{}

\renewcommand{\shorttitle}{}

\begin{document}
\maketitle

\begin{abstract}
Extreme precipitation events occurring over large spatial domains pose substantial threats to societies because they can trigger compound flooding, landslides, and infrastructure failures across wide areas. A hybrid framework for spatial extreme precipitation modeling and risk zoning is proposed that integrates graph neural networks with $r$-Pareto processes (GNN-$r$P). Unlike traditional statistical spatial extremes models, this approach learns nonlinear, nonstationary dependence structures from precipitation-derived spatial graphs and applies a data-driven tail functional to model joint exceedances in a low-dimensional embedding space. Using NASA’s IMERG observations (2000-2021) and CMIP6 SSP5-8.5 projections, the framework delineates coherent high-risk zones, quantifies their temporal persistence, and detects emerging hotspots under climate change. Compared with two baseline approaches, the GNN-$r$P pipeline substantially improves pointwise detection of high-risk grid cells while yielding comparable clustering stability. Results highlight persistent high-risk regions in the tropical belt, especially monsoon and convective zones, and reveal decadal-scale persistence that is punctuated by episodic reconfigurations under high-emission scenarios. By coupling machine learning with extreme value theory, GNN-$r$P offers a scalable, interpretable tool for adaptive climate risk zoning, with direct applications in infrastructure planning, disaster preparedness, and climate-resilient policy design.
\end{abstract}


\section{Introduction}
Extreme precipitation events over large spatial domains are key drivers of hydrological disasters such as floods, landslides, and infrastructure failures. These risks are particularly acute in tropical and subtropical regions, including the Indo-Pacific, equatorial Africa, and northern South America, where both precipitation intensity and population vulnerability are high. The compound nature of these events, often characterized by prolonged heavy rainfall followed by rapid runoff, can overwhelm natural and engineered drainage systems, triggering cascading hazards across urban and rural landscapes. So comparable meteorological extremes often produce disproportionately larger mortality, long-term livelihoods disruption, and economic loss in these settings \parencite{bouwer2011,doocy2013,RN11}. Beyond immediate impacts, recurrent extremes drive persistent secondary effects, including soil degradation, agricultural yield shocks, damaged transport and energy networks, and increased displacement, which in aggregate raise the social and fiscal costs of recovery and adaptation. Such extremes contribute substantially to mortality and economic losses, drawing attention from international bodies including the United Nations and the Intergovernmental Panel on Climate Change (IPCC).

Accurate spatial modeling of precipitation extremes is essential for climate-resilient infrastructure, risk-based land-use planning, and operational disaster mitigation. Classical spatial extremes approaches, including max-stable processes and threshold-based generalized Pareto constructions, supply rigorous asymptotic theory for tail behaviour and uncertainty quantification. But they commonly rely on asymptotic, stationarity, and parametric assumptions that limit flexibility when applied to high-dimensional gridded datasets exhibiting nonstationarity, localized dependence, or transitions between asymptotic dependence and independence \parencite{davison_statistics_2015-1,huser_advances_2022}, that limit their ability to capture complex dependence in large-scale gridded datasets \autocite{engelke_graphical_2019}. Widely used climate products such as ERA5\footnote{ERA5: European Centre for Medium-Range Weather Forecasts (ECMWF) Fifth Generation Reanalysis}, ERA5-Land\footnote{ERA5-Land: European Centre for Medium-Range Weather Forecasts (ECMWF) Fifth Generation Reanalysis - Land}, MERRA-2\footnote{MERRA-2: Modern-Era Retrospective analysis for Research and Applications, Version 2}, PRISM\footnote{PRISM: Parameter-elevation Relationships on Independent Slopes Model}, which provide valuable inputs but often misrepresent the magnitude and spatial structure of extremes, especially in mountainous, coastal, and poorly instrumented regions \parencite{smith_evaluating_2025}. These limitations propagate into modelled risk estimates and infrastructure design loads unless corrected for, and they increase epistemic uncertainty when projecting future changes under warming scenarios, particularly because temperature-precipitation sensitivities and the clustering behaviour of extremes may evolve with climate change \parencite{WASKO2017575}.

The $r$-Pareto process \autocite{ferreira_generalized_2014,dombry_pareto_2024} extends threshold-based approaches by modeling joint exceedances via user-defined spatial risk functionals (e.g., regional mean, maximum tail intensity, domain-total precipitation), enabling richer characterizations of compound events. While recent work has improved estimation in high-dimensional, nonstationary settings via score matching \autocite{de_fondeville_high-dimensional_2018,lederer_extremes_2024} and climate-covariate extensions \autocite{zhong_spatial_2025}, most methods still require fixed functionals and dependence structures. To address these limitations, machine-learning-based spatial models have emerged, with graph neural networks (GNNs) particularly suited to irregular, nonstationary domains. GNNs represent locations as nodes linked by geographic or statistical relationships, capturing multiscale dependencies without requiring Euclidean grids. Architectures such as Graph Sample and Aggregate (GraphSAGE) \autocite{hamilton_inductive_2017} and Neural Network-Generalized Least Squares (NN-GLS) \autocite{zhan_neural_2025} have been successfully applied in precipitation forecasting, compound-extreme detection, and ensemble post-processing \autocite{lam_graphcast_2023,koh_using_2024,bulte_graph_2025}.

This study proposes a hybrid GNN with $r$-Pareto (GNN-$r$P) framework that learns low-dimensional spatial embeddings via a GNN autoencoder. The GNN-rP framework defines a data-driven tail functional from embeddings produced by an autoencoder of GNN and combines it with the $r$-Pareto process to model joint exceedances. The framework captures nonlinear, high-dimensional, and nonstationary dependence without relying on fixed kernels or handcrafted functionals. It also clusters embeddings to delineate coherent risk zones. We benchmark the GNN-$r$P framework against two tail-based zoning strategies, applying all methods to IMERG precipitation data (2000-2021) and CMIP6 SSP5-8.5 projections. Temporal consistency is evaluated using the Adjusted Rand Index and precision-recall metrics. Results show that GNN-$r$P improves spatial coherence and stability in high-risk zone detection, with particularly strong performance in the tropical belt and monsoon regions. It also identifies emerging hotspots under climate change. By integrating deep learning with extreme value theory, this framework provides a principled and scalable approach for adaptive spatial risk zoning, supporting climate-resilient planning and early warning systems.

This paper makes three primary contributions. First, a hybrid framework that integrates graph neural network (GNN) embeddings with the $r$-Pareto tail process is developed. This framework enables the learning of data-driven spatial risk functionals and the modeling of joint exceedances in a low-dimensional latent space. Unlike fixed, handcrafted spatial kernels, it uses learned node embeddings to define tail severity and performs $r$-Pareto normalization in the embedding domain, thereby enhancing flexibility in capturing nonstationary and nonlinear tail dependence. Second, at the algorithmic level, a stable estimation pipeline that combines a GNN autoencoder with $r$-Pareto feature construction and clustering is introduced. This pipeline improves numerical robustness when learning tail dependence in high-dimensional embedding spaces. Finally, an extensive empirical evaluation is conducted using IMERG observations (2000-2021) and bias-corrected CMIP6 SSP5-8.5 projections. This evaluation includes global risk zoning, cross-year Mahalanobis diagnostics, and systematic comparisons against two baseline tail-based zoning methods using precision, recall, and adjusted Rand Index metrics. Additionally, case studies further demonstrate the method’s practical value by identifying persistent tropical hotspots and emerging future risk regions. The implications of this method for infrastructure planning, early warning systems, and climate-sensitive insurance design are also discussed \parencite{huser_advances_2022,koh_using_2024,RN11}.

To clarify the study’s logic and structure, the subsequent content unfolds as follows. Section~\ref{sec:empirical} introduces two core datasets. One is NASA’s IMERG precipitation observations (2000-2021), which have high spatiotemporal resolution and are used for historical analysis. The other is bias-corrected CMIP6 SSP5-8.5 projections, applied to simulate future scenarios. This section also explains methods for precipitation feature construction and spatial graph normalization. Next, Section~\ref{sec:model} systematically elaborates on the technical workflow of the GNN-$r$P hybrid framework. It covers spatial graph construction, graph convolutional embedding learning, embedding-based tail functional design, $r$-Pareto normalization, and augmented embedding clustering for risk zoning. Then, Section~\ref{sec:calibration} specifies the GNN architecture, training parameters, and step-by-step computational procedures, which are summarized in Algorithm \ref{alg:gnn_rpareto}. Following this, Section~\ref{sec:riskzoning} validates the framework’s effectiveness. It does so through global risk zoning, cross-year Mahalanobis distance analysis, comparisons with baseline models (which confirm the GNN-$r$P’s superior precision and recall), and temporal consistency analysis under the SSP5-8.5 scenario. Finally, Section~\ref{sec:conclusion} synthesizes key findings, highlights practical implications such as applications for climate adaptation and risk financing, notes limitations, and proposes future extensions including dynamic graphs and multi-hazard coupling.

\section{Data} \label{sec:empirical}

To assess the performance and practical utility of the proposed GNN-$r$P framework, an empirical study based on both satellite observations and climate model simulations is constructed. The analysis is designed to capture the spatial structure and temporal dynamics of extreme precipitation, leveraging learned graph embeddings that encode neighborhood dependence and variability in the upper tail. Two complementary sources of gridded precipitation data are employed:
\begin{itemize}
	\item \textbf{IMERG (Integrated Multi-satellitE Retrievals for GPM)}: A satellite-based global precipitation product developed by NASA as part of the Global Precipitation Measurement (GPM) mission. IMERG provides near-global coverage (60°S-60°N) with high spatial ($0.1^\circ$) and temporal (30-minute) resolution. We use the Version 6.0, which incorporates gauge-adjusted estimates to improve reliability. IMERG is particularly well-suited for analyzing spatial extremes due to its fine resolution, consistent global coverage, and proven performance in tropical and subtropical regions where ground-based data are sparse. However, at the time of writing, IMERG Version 6.0 is only available through December 2021. This limitation constrains our historical analysis period but is offset by the dataset’s high temporal fidelity and spatial resolution compared to reanalysis alternatives.
	
	\item \textbf{CMIP6 (SSP5-8.5 scenario)}: Daily precipitation simulations are sourced from the Coupled Model Intercomparison Project Phase 6 (CMIP6), specifically under the Shared Socioeconomic Pathway 5-8.5 (SSP5-8.5) scenario. This high-emission scenario assumes continued fossil-fuel intensive development and represents a pessimistic, yet policy-relevant pathway for long-term climate risk planning.
	
	SSP5-8.5 is widely used in climate impact studies as a stress-test scenario, helping to identify regions that may experience amplified risks under unmitigated global warming. It is especially appropriate for modeling extremes because it exhibits stronger projected changes in precipitation intensity, variability, and tail behavior relative to other scenarios. These amplified signals facilitate clearer evaluation of model performance and provide conservative projections for infrastructure and policy resilience studies.
	
	CMIP6 model outputs are selected based on data availability, completeness, and spatial resolution. All simulations are bias-corrected and regridded to align with the IMERG spatial structure, enabling direct comparison of historical and future risk zones.
\end{itemize}

\section{Methodology} \label{sec:model}
This section introduces the proposed GNN-$r$P framework, which integrates graph-based representation learning with extreme value theory to characterize spatial extremes of precipitation. The methodology proceeds in a sequence of components: spatial graph construction, graph convolutional embedding, tail functional definition, $r$-Pareto normalization and feature construction, clustering-based risk zoning, and cross-year Mahalanobis distance analysis for temporal change detection. Together, these steps enable the extraction of nonlinear spatial dependencies, identification of extremes, quantification of their dynamics, and generation of coherent risk zones across both historical observations and climate projections.

\subsection{Spatial graph construction}
We assume that a preprocessed and standardized feature matrix of precipitation
\[
\mathbf{X} \;=\; \begin{bmatrix} \mathbf{x}_1^\top \\ \vdots \\ \mathbf{x}_N^\top \end{bmatrix}
\in\mathbb{R}^{N\times d_{\mathrm{in}}}
\]
is available for a given year, where \(N\) is the number of retained spatial grid cells, \(\mathbf{x}_i\in\mathbb{R}^{d_{\mathrm{in}}}\) is the feature vector at node \(i\), and \(d_{\mathrm{in}}\) denotes the input feature dimension. Denote by \(\mathcal{G}=(\mathcal{V},\mathcal{E})\) the undirected spatial graph with node set \(\mathcal{V}=\{1,\dots,N\}\) and edge set \(\mathcal{E}\). The adjacency matrix of \(\mathcal{G}\) is \(\mathbf{A}\in\{0,1\}^{N\times N}\).

The spatial graph \(\mathcal{G}\) is constructed by connecting each node to its \(k\) nearest neighbors in geographic Euclidean distance, producing the edge set \(\mathcal{E}\) and adjacency matrix \(\mathbf{A}\). The graph is fixed for the year under analysis and encodes local spatial topology used by the graph convolutional encoder.

\subsection{Graph convolutional embedding} \label{sec:embedding}
Let \(\mathbf{H}^{(0)}=\mathbf{X}\). For a graph convolutional encoder with layer index \(l\), node feature matrix \(\mathbf{H}^{(l)}\in\mathbb{R}^{N\times d^{(l)}}\), and weight matrix \(\mathbf{W}^{(l)}\), a standard propagation rule can be written as
\[
\mathbf{H}^{(l+1)} \;=\; \sigma\!\big(\widetilde{\mathbf{D}}^{-1/2}\widetilde{\mathbf{A}}\widetilde{\mathbf{D}}^{-1/2}\,\mathbf{H}^{(l)}\mathbf{W}^{(l)}\big),
\]
where \(\widetilde{\mathbf{A}}=\mathbf{A}+\mathbf{I}_N\), \(\widetilde{\mathbf{D}}=\operatorname{diag}(\widetilde{\mathbf{A}}\mathbf{1})\), and \(\sigma(\cdot)\) is an elementwise nonlinearity (e.g., ReLU, distinct from GPD scale parameter \(\sigma\) below). A two-layer encoder of this form defines a parametric map \(f_{\theta}\) that yields the latent embedding matrix
\[
\mathbf{Z} \;=\; f_{\theta}(\mathbf{X},\mathcal{E}) \;=\; \begin{bmatrix} \mathbf{z}_1^\top \\ \vdots \\ \mathbf{z}_N^\top \end{bmatrix}
\in\mathbb{R}^{N\times d_z},
\]
where \(d_z\) is the latent dimension and \(\mathbf{z}_i\in\mathbb{R}^{d_z}\) denotes the embedding for node \(i\). This class of graph convolutional architectures is standard in the literature on representation learning for graphs \parencite{kipf_semi-supervised_2017,hamilton_inductive_2017}. A decoder \(g_{\phi}\) (parameter \(\phi\) distinct from encoder's \(\theta\)) maps \(\mathbf{Z}\) back to reconstructed features \(\widehat{\mathbf{X}}=g_{\phi}(\mathbf{Z})\), and the encoder and decoder are trained jointly by minimizing the reconstruction loss
\[
\mathcal{L}_{\mathrm{GNN}}(\theta,\phi) \;=\; \frac{1}{N}\sum_{i=1}^N \\ ||\mathbf{x}_i - g_{\phi}(\mathbf{z}_i)\\||_2^2,
\]
so that the learned \(\mathbf{Z}\) captures nonlinear and multiscale spatial dependencies present in \(\mathbf{X}\).

\subsection{Embedding-based Tail Functional and $r$-Pareto Feature Construction}
Define an embedding-based scalar risk functional\label{sec:tail} \(r:\mathbb{R}^{d_z}\to\mathbb{R}\) that maps each latent vector to a tail severity score. In this work we use the componentwise mean:
\[
r_i \;=\; r(\mathbf{z}_i) \;=\; \frac{1}{d_z}\sum_{j=1}^{d_z} z_{ij},\qquad i=1,\dots,N,
\]
where \(z_{ij}\) denotes the \(j\)-th component of \(\mathbf{z}_i\). For a chosen tail quantile \(\tau\in(0,1)\) (for example \(\tau=0.95\)), let \(q_{\tau}(r)\) denote the empirical \(\tau\)-quantile of \(\{r_i\}_{i=1}^N\). The tail index set is
\[
\mathcal{I}^* \;=\; \{\, i\in\mathcal{V} : r_i \ge q_{\tau}(r)\,\}.
\]

For $i\in\mathcal{I}^*$ denote the tail values by $r_i$ and form nonnegative excesses
\[
\tilde r_i = r_i - \min_{j\in\mathcal{I}^*} r_j, \qquad i\in\mathcal{I}^*.
\]

A generalized Pareto distribution (GPD) is fitted to $\{\tilde r_i\}$ with shape $\xi$, scale $\sigma>0$ (distinct from activation function $\sigma(\cdot)$), and location $\mu$, using the cumulative distribution function $F_{\mathrm{GPD}}(\cdot; \xi, \sigma, \mu)$. Denote fitted parameters by $(\hat{\xi}, \hat{\sigma}, \hat{\mu})$.

For each tail node $i \in \mathcal{I}^*$, define the four-dimensional $r$-Pareto feature vector $\mathbf{R}_i \in \mathbb{R}^4$ as
\[
\begin{aligned}
	R_{i,1} &= F_{\mathrm{GPD}}(r_i; \hat{\xi}, \hat{\sigma}, \hat{\mu}),\\
	R_{i,2} &= \frac{\tilde r_i}{\hat{\sigma}},\\
	R_{i,3} &= \hat{\xi},\\
	R_{i,4} &= \hat{\sigma}.
\end{aligned}
\]

For non-tail nodes $i \notin \mathcal{I}^*$, set $\mathbf{R}_i = \mathbf{0}$. Stack all $\mathbf{R}_i$ to form the matrix
\[
\mathbf{R} = \begin{bmatrix} \mathbf{R}_1^\top \\ \vdots \\ \mathbf{R}_N^\top \end{bmatrix} \in \mathbb{R}^{N \times 4}.
\] The first two components provide a normalized, probabilistic measure of extremeness and comparable exceedance magnitude while the last two record the fitted tail parameters used for normalization and interpretation. The use of a functional-defined exceedance set and GPD-based exceedance parametrization is consistent with recent high-dimensional POT methodology \parencite{de_fondeville_high-dimensional_2018,ferreira_generalized_2014}.

\subsection{Augmented embedding and risk zoning}
Risk zoning is performed by applying \(K\)-means clustering to the augmented embeddings \(\mathbf{Z}_{\mathrm{aug}}\). This procedure groups spatial locations that share similar latent features and tail signatures into a predetermined number of zones, thereby producing coherent risk regions for interpretation and decision-making.

Form the augmented embedding
\[
\mathbf{Z}_{\mathrm{aug}} \;=\; \big[\,\mathbf{Z}\ \ \mathbf{R}\,\big] \in\mathbb{R}^{N\times(d_z+4)}.
\]

Let \(K_c\) be the desired number of risk zones and let \(\boldsymbol{\mu}_1,\dots,\boldsymbol{\mu}_{K_c}\in\mathbb{R}^{d_z+4}\) denote cluster centroids. A clustering partition \(\{c_i\}_{i=1}^N\) with labels \(c_i\in\{1,\dots,K_c\}\) may be characterized as a minimizer of the within-cluster sum of squared distances
\[
J(\{c_i\},\{\boldsymbol{\mu}_k\}) \;=\; \sum_{i=1}^N \big\lVert \mathbf{z}_{\mathrm{aug},i} - \boldsymbol{\mu}_{c_i}\big\rVert_2^2,
\]
where \(\mathbf{z}_{\mathrm{aug},i}\) denotes the \(i\)-th row of \(\mathbf{Z}_{\mathrm{aug}}\). Because the matrix \(\mathbf{R}\) injects tail-specific information into \(\mathbf{Z}_{\mathrm{aug}}\), the resulting clusters emphasize joint exceedance behavior together with latent directional similarity rather than only pointwise tail magnitude. The centroid-based clustering objective above corresponds to the classical \(K\)-means formulation \parencite{macqueen_some_1967}.

In practice, clustering is applied to the full \(\mathbf{Z}_{\mathrm{aug}}\) so that non-tail nodes with \(\mathbf{R}_i=\mathbf{0}\) are assigned to clusters whose centroids reflect both latent geometry and tail signatures. As an alternative assignment strategy one can propagate labels from tail nodes by assigning a non-tail node \(j\) the label of its nearest tail node in latent space,
\[
c_j \;=\; c_{i^\star}, \quad\text{where } i^\star=\arg\min_{i\in\mathcal{I}^*}\big\lVert\mathbf{z}_j-\mathbf{z}_i\big\rVert_2.
\]
Either strategy yields a complete spatial risk zoning map. Selecting between them depends on the analyst's preference for tail-driven propagation or centroid-driven assignment and on the intended use of the zoning output.

\section{Model Calibration}\label{sec:calibration}
\subsection{Feature Construction, Graph Formation, and Normalization}\label{subsec:feature_graph_norm}

For a given year $y$, let each spatial node $i \in \mathcal{V}=\{1,\dots,N\}$ be represented by a feature vector
\[
\mathbf{x}_i^{(y)} = \begin{bmatrix} \mu_i^{(y)}, \sigma_i^{(y)}, M_i^{(y)} , f_i^{(y)} \end{bmatrix}^\top \in \mathbb{R}^{d_{\mathrm{in}}},
\]
where $\mu_i^{(y)}$, $\sigma_i^{(y)}$, $M_i^{(y)}$, and $f_i^{(y)}$ denote mean, standard deviation, maximum, and threshold-exceedance frequency of daily precipitation ${P_{i,t}^{(y)}}_{t=1}^{T_y}$ at node $i$ respectively, where  $T_y$ represents the number of daily observations within year $y$.

\[
\begin{aligned}
	\mu_i^{(y)} &= \frac{1}{T_y} \sum_{t=1}^{T_y} P_{i,t}^{(y)} 
	&& \text{(mean precipitation)}, \\[0.5em]
	\sigma_i^{(y)} &= \sqrt{ \frac{1}{T_y - 1} \sum_{t=1}^{T_y} \left( P_{i,t}^{(y)} - \mu_i^{(y)} \right)^2 }
	&& \text{(standard deviation)}, \\[0.5em]
	M_i^{(y)} &= \max_{1 \le t \le T_y} P_{i,t}^{(y)}
	&& \text{(maximum daily precipitation)}, \\[0.5em]
	f_i^{(y)} &= \frac{1}{T_y} \sum_{t=1}^{T_y} \mathbf{1} \left\{ P_{i,t}^{(y)} > q_{\tau}^{(y)} \right\}
	&& \text{(frequency of extremes)}.
\end{aligned}
\]

Stacking all node features produces the feature matrix
\[
\mathbf{X}^{(y)} = \begin{bmatrix} (\mathbf{x}_1^{(y)})^\top \\ \vdots \\ (\mathbf{x}_N^{(y)})^\top \end{bmatrix} \in \mathbb{R}^{N \times d_{\mathrm{in}}}.
\]

A spatial graph $\mathcal{G}=(\mathcal{V},\mathcal{E})$ is then constructed by connecting each node to its $k$ nearest neighbors in geographic space. Features are standardized across nodes to zero mean and unit variance per dimension to form the normalized matrix \(\mathbf{X}^{(y)}\).

\subsection{Model Implementation} \label{sec:alg}
In line with the algorithmic description in Section~\ref{sec:alg}, we implement the spatial risk zoning pipeline as follows. For each year, node-level features are computed from the preprocessed IMERG precipitation data and linked into a $k=8$ nearest-neighbor spatial graph. The trained GCN encoder produces low-dimensional embeddings $\mathbf{Z}^{(y)} \in \mathbb{R}^{N \times d_z}$, from which tail scores $r_i^{(y)}$ are calculated. Extreme nodes are defined as those exceeding the $95$th percentile of $r_i^{(y)}$ values. Their embeddings are normalized by $r$-Pareto scaling \autocite{de_fondeville_high-dimensional_2018}, yielding direction-preserving vectors $\mathbf{y}_i^{(y)} = \mathbf{z}_i^{(y)} / r_i^{(y)}$. We then perform $K$-means clustering on $\mathbf{Z}^{(y)}$ to obtain annual spatial risk zones. Nodes without extreme classification inherit the label of the nearest extreme node in embedding space, ensuring a complete risk zoning map for all grid cells. This procedure closely follows the conceptual framework proposed by \textcite{de_fondeville_high-dimensional_2018}, and we obtain normalized tail embeddings via the $r$-Pareto normalization, construct the $r$-Pareto feature vectors $\mathbf{R}$, augment the latent embeddings and then derive spatial risk zones through clustering. The step-by-step computational process is summarized in Algorithm~\ref{alg:gnn_rpareto}.

\begin{algorithm}[hbt!]
	\caption{GNN-$r$P Framework for Spatial Extreme Precipitation Modeling}
	\label{alg:gnn_rpareto}
	
	\KwIn{
		Gridded precipitation data $\{P^{(y)}\}$ for years $y = t_{1}, \ldots, {t_{n}}$;\\
		Quantile threshold $\tau \in (0, 1)$ (e.g., $\tau = 0.95$);\\
		GNN architecture parameters: input dim $d_{\mathrm{in}}$, hidden dim $d_h$, latent dim $d_z$;\\
		Number of training epochs $T_{\mathrm{train}}$;\\
		Number of neighbors $k$ for KD-tree graph;\\
		Number of clusters $K_c$ for risk zoning.
	}
	
	\KwOut{
		Node embeddings $\mathbf{Z}^{(y)}$ for each year;\\
		Tail-normalized extreme vectors $\mathbf{Y}^{(y)} = \{\mathbf{y}_i^{(y)}\}_{i \in \mathcal{I}^*}$;\\
		Risk zone labels $\{c_i^{(y)}\}$.
	}
	
	\ForEach{year $y \in [t_{1}, t_{n}]$}{
		\tcp{Step 1: Feature Extraction}
		Load gridded precipitation data $P^{(y)}$\;
		Resample to daily resolution if needed\;
		For each spatial location $i$, compute:\\
		\quad mean $\mu_i^{(y)}$, std $\sigma_i^{(y)}$, max $M_i^{(y)}$, threshold freq. $f_i^{(y)}$ above $q^{(y)}_{\tau}$\;
		Construct feature vector $\mathbf{x}_i^{(y)} = [\mu_i^{(y)}, \sigma_i^{(y)}, M_i^{(y)}, f_i^{(y)}] \in \mathbb{R}^{d_{\mathrm{in}}}$\;
		Standardize features across all valid nodes\;
		
		\tcp{Step 2: Spatial Graph Construction}
		Build KD-tree over spatial locations\;
		Connect each node to $K$ nearest neighbors to form edge set $\mathcal{E}$\;
		
		\tcp{Step 3: GNN Embedding Learning}
		Train GCN encoder $f_\theta$ and decoder $g_\phi$ on $\mathbf{x}_i^{(y)}$ for $T_{\mathrm{train}}$ epochs\;
		Minimize autoencoder loss:\\
		\quad $\mathcal{L}_{\mathrm{GNN}}(\theta,\phi) = \frac{1}{N}\sum_i \ \ ||\mathbf{x}_i^{(y)} - g_\phi(f_\theta(\mathbf{x}_i^{(y)}))\ ||_2^2$\;
		Obtain embeddings $\mathbf{Z}^{(y)} = \{\mathbf{z}_i^{(y)} \in \mathbb{R}^{d_z}\}$\;
		
		\tcp{Step 4: Tail Extraction and Normalization}
		Compute tail score $r_i^{(y)} = \frac{1}{d_z} \sum_{j=1}^{d_z} z_{ij}^{(y)}$\;
		Define tail set $\mathcal{I}^{*(y)} = \{i : r_i^{(y)} > \text{quantile}_\tau(\{r_i^{(y)}\})\}$\;
		Normalize: $\mathbf{y}_i^{(y)} = \mathbf{z}_i^{(y)} / r_i^{(y)}$, for $i \in \mathcal{I}^{*(y)}$\;
				
		\tcp{Step 5: Risk Zone Clustering}
		Apply $K$-means clustering on $\mathbf{Z}^{(y)}$ with $K_c$ clusters\;
		Assign zone label $c_i^{(y)}$ to each node $i$\;
		
		\tcp{Step 6: Output Storage}
		Save: $\mathbf{Z}^{(y)}$, $\{\mathbf{y}_i^{(y)}\}_{i \in \mathcal{I}^{*(y)}}$, $\{c_i^{(y)}\}$, and spatial index list\;
	}
\end{algorithm}

\section{Empirical Study} \label{sec:riskzoning}
Building on the methodological foundation outlined in Section~\ref{sec:model}, this section clarifies the two main application scenarios of the proposed pipeline and evaluates its effectiveness through empirical analysis.

The pipeline is designed for two distinct purposes. First, when the GNN encoder is trained independently on observational data from each year, the framework produces retrospective annual risk zonations, characterizing historical spatial patterns of extreme precipitation risk and their interannual variability. Second, under scenario-based projection with CMIP6 (SSP5-8.5), the same pipeline is applied to bias-corrected and regridded climate model outputs, yielding scenario-driven projections rather than deterministic forecasts.

The subsequent empirical analysis focuses on four objectives: (i) capturing spatiotemporal patterns of extreme precipitation risk, (ii) quantifying dynamic changes in risk zones, (iii) benchmarking against competing methods, and (iv) adapting the framework to future climate scenarios. Together, these results provide empirical support for spatial delineation and temporal evolution analysis of extreme precipitation risk.

\subsection{Model Configuration and Training}
\label{sec:training}

The graph convolutional autoencoder described in Section~\ref{sec:model} is trained independently for each year. The encoder consists of two graph convolutional layers, each followed by a nonlinear activation, mapping standardized node-level features into a latent representation of dimension $d_z$. The decoder is a linear transformation that reconstructs the original input feature vectors from the embeddings. Formally, given input $\mathbf{x}_i \in \mathbb{R}^{d_{\mathrm{in}}}$ at node $i$, the encoder produces an embedding $\mathbf{z}_i \in \mathbb{R}^{d_z}$, and the decoder reconstructs $\hat{\mathbf{x}}_i \in \mathbb{R}^{d_{\mathrm{in}}}$.

Training minimizes the reconstruction loss
\[
\mathcal{L}_{\mathrm{GNN}}(\theta,\phi) \;=\; \frac{1}{N}\sum_{i=1}^N \\|| \mathbf{x}_i - \hat{\mathbf{x}}_i \\||_2^2,
\]

where $\hat{\mathbf{x}}_i = g_\phi(f_\theta(\mathbf{x}_i))$ is the decoder output, and $(\theta,\phi)$ denote the encoder and decoder parameters, respectively. This ensures that the learned embeddings $\mathbf{z}_i$ preserve essential spatial and distributional structure contained in the input features.

Spatial graphs are constructed using $k=8$ nearest neighbors in Euclidean distance between node coordinates, consistent across all years. After training, tail scores $r_i = r(\mathbf{z}_i)$ are computed as in Section~\ref{sec:tail}, and nodes exceeding the $\tau$-quantile threshold (with $\tau=0.95$ in our implementation) are identified as extremes. These nodes are then normalized under the $r$-Pareto scheme and used to construct the augmented embedding $\mathbf{Z}_{\mathrm{aug}}$ for clustering-based risk zoning in Section~\ref{subsec:zonation}.

We apply this framework to NASA’s GPM IMERG Version~07 Final Run dataset, covering the years 2000-2021. The data are aggregated to a $0.5^\circ \times 0.5^\circ$ latitude-longitude grid, from which annual feature matrices $\mathbf{X}^{(y)}$ and spatial graphs $\mathcal{G}^{(y)}$ are constructed. This results in a consistent set of annual embeddings $\mathbf{Z}^{(y)}$ and corresponding risk zoning outputs across the study period.

\subsection{Global Risk Zonation}\label{subsec:zonation}
\begin{figure*}
	\centering
	\subfloat[most common risk zone classification]{
		\includegraphics[width=0.5\linewidth]{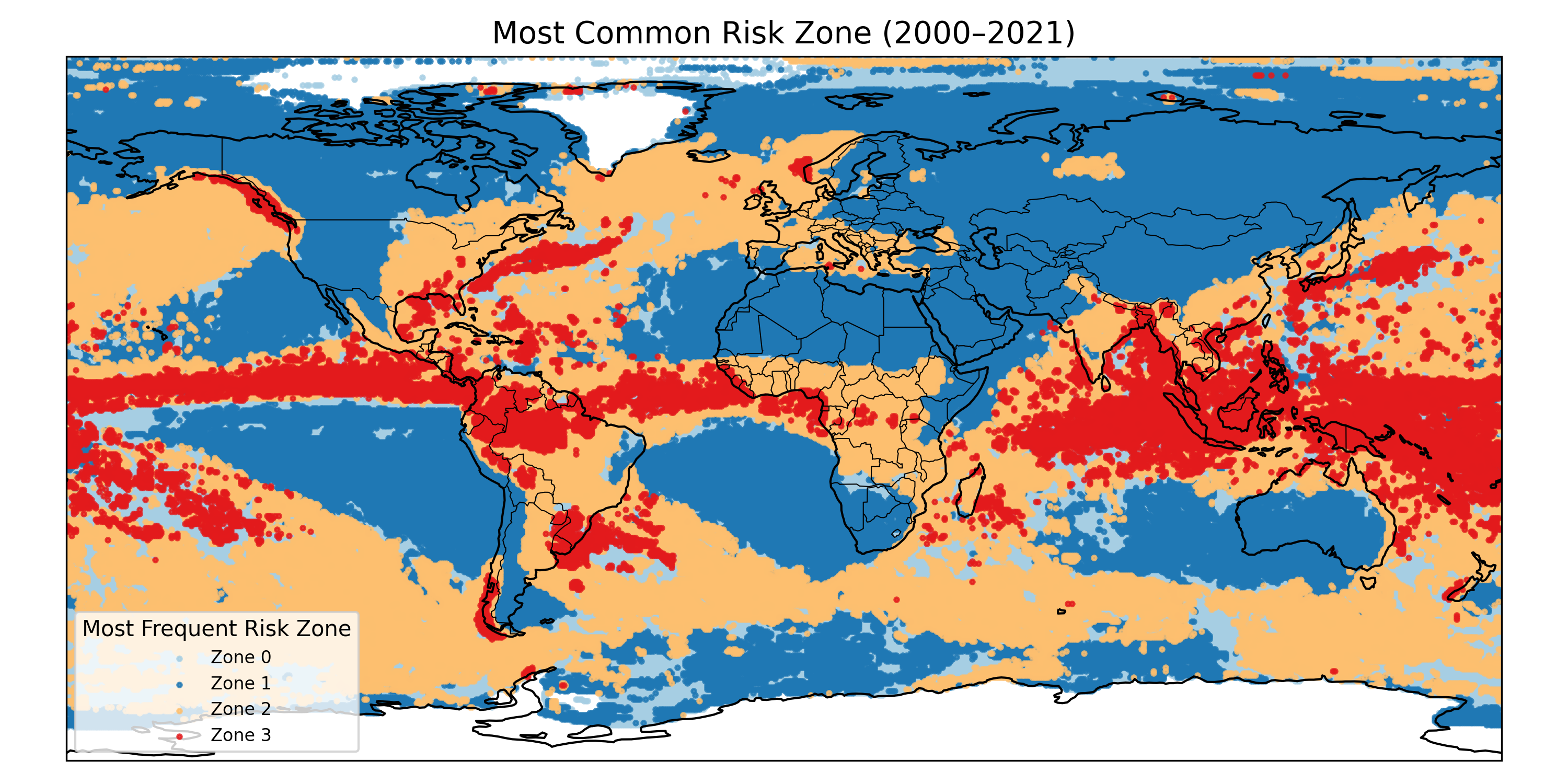}
		\label{fig:riskzonemodemap}
	}
	\subfloat[high-risk zone frequency]{
		\includegraphics[width=0.5\linewidth]{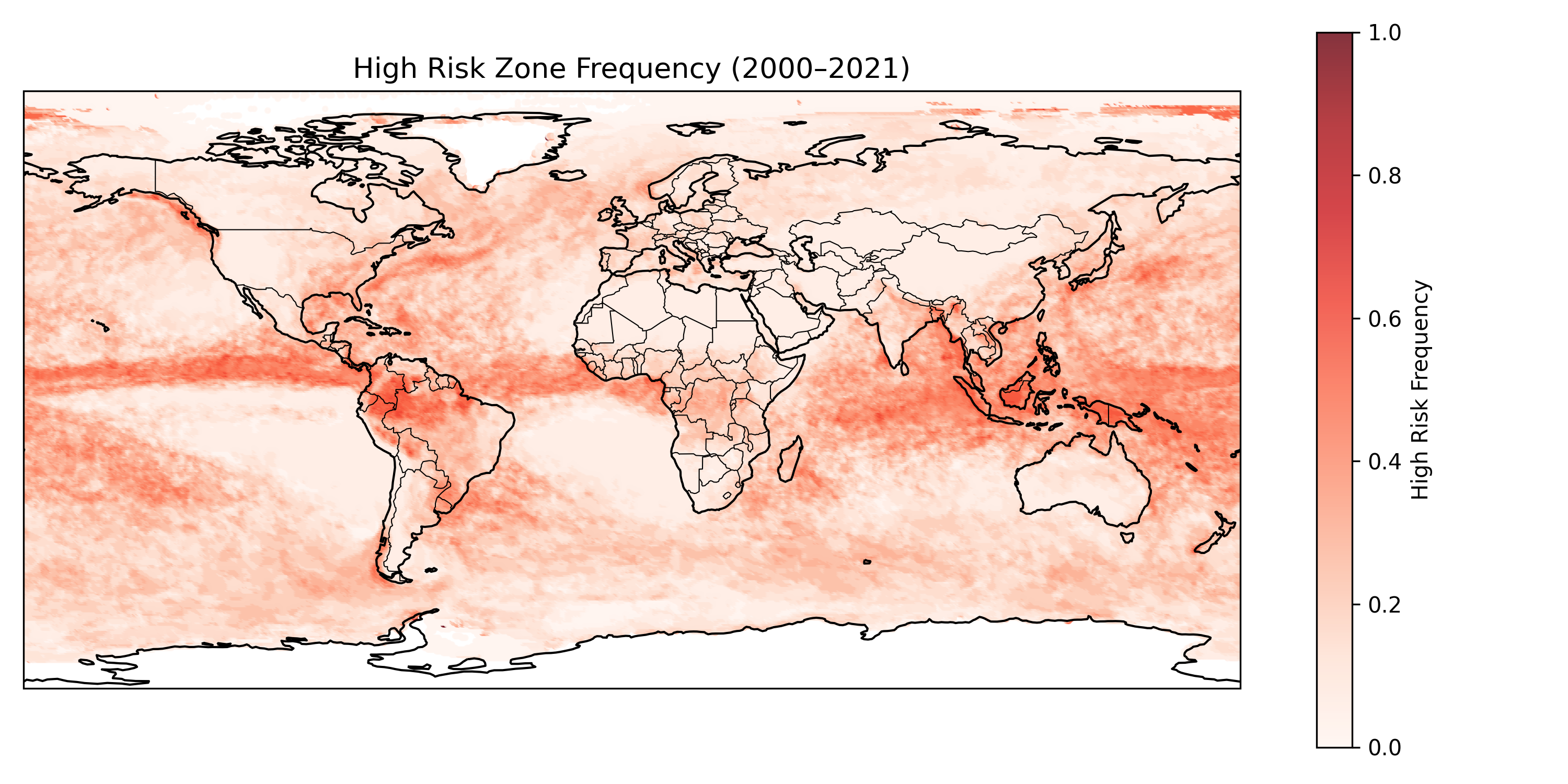}
		\label{fig:riskzonestabilitymap}
	}
	\caption{Global map of high-risk zone produced by the GNN-$r$P model.}
	\label{fig:my_label1}
\end{figure*}

Following the spatial risk zoning procedure described in Section~\ref{sec:alg}, each grid cell in Figure~\ref{fig:riskzonemodemap} is colored by the modal risk zone (0-3) observed over all years. Blue indicates the lowest-risk zone 0 and red indicates the highest-risk zone 3. The modal classification at a given location is obtained by tallying its annual zone labels across the 2000-2021 record and selecting the most frequently occurring label. This choice emphasizes persistent exposure patterns and reduces sensitivity to transient single-year anomalies. The figure shows that tropical regions consistently fall into higher-risk categories, shown in red and orange, while polar and desert regions are predominantly low risk and appear in blue. 

The global pattern in Figure~\ref{fig:riskzonemodemap} reveals a clear contrast between tropical and extratropical regions. Equatorial belts such as the Amazon, the Congo and the Maritime Continent are mostly classified as high risk. By contrast, mid and high latitude continents including North America, Eurasia and Antarctica remain in low or moderate risk zones. Regionally, coastal monsoon and hurricane-prone areas such as Southeast Asia and the Gulf Coast tend to appear in higher risk classes, whereas interior arid or polar regions largely remain in zone 0 or zone 1. This spatial contrast agrees with climatological evidence showing the highest intensity and frequency of extreme precipitation events in the tropics, especially in convectively active regions \autocite{trenberth_relationships_2005, hersbach_era5_2020}.

It is important to note that the modal map summarizes long-term persistence and does not capture year-to-year variability. Locations that change label frequently across years are still assigned a single modal class in this map. For a complementary view of temporal stability we present frequency-based diagnostics in Figure~\ref{fig:riskzonestabilitymap}. Those diagnostics quantify how often each cell is classified into the highest-risk category and thereby reveal regions of consistent exposure as well as regions of episodic exposure.

Colors in the frequency map denote the proportion of years each location was classified as high risk, with darker red indicating higher frequency. The map in Figure~\ref{fig:riskzonestabilitymap} shows that certain tropical areas are repeatedly flagged as high risk in most years. Notably, the Indo-Pacific warm pool, tropical Africa and parts of South America exhibit frequencies above 0.6. By contrast, nearly all mid-latitude and polar regions have frequencies below 0.2, which indicates that high precipitation extremes occur rarely in those areas. These patterns reinforce the view that extreme precipitation risk is concentrated in the tropics and equatorial oceans where convective and monsoon processes are strongest \autocite{masson-delmotte_summary_2021, schumacher_organization_2005}.

\begin{figure}[hbt!]
	\centering
	\includegraphics[width=0.75\linewidth]{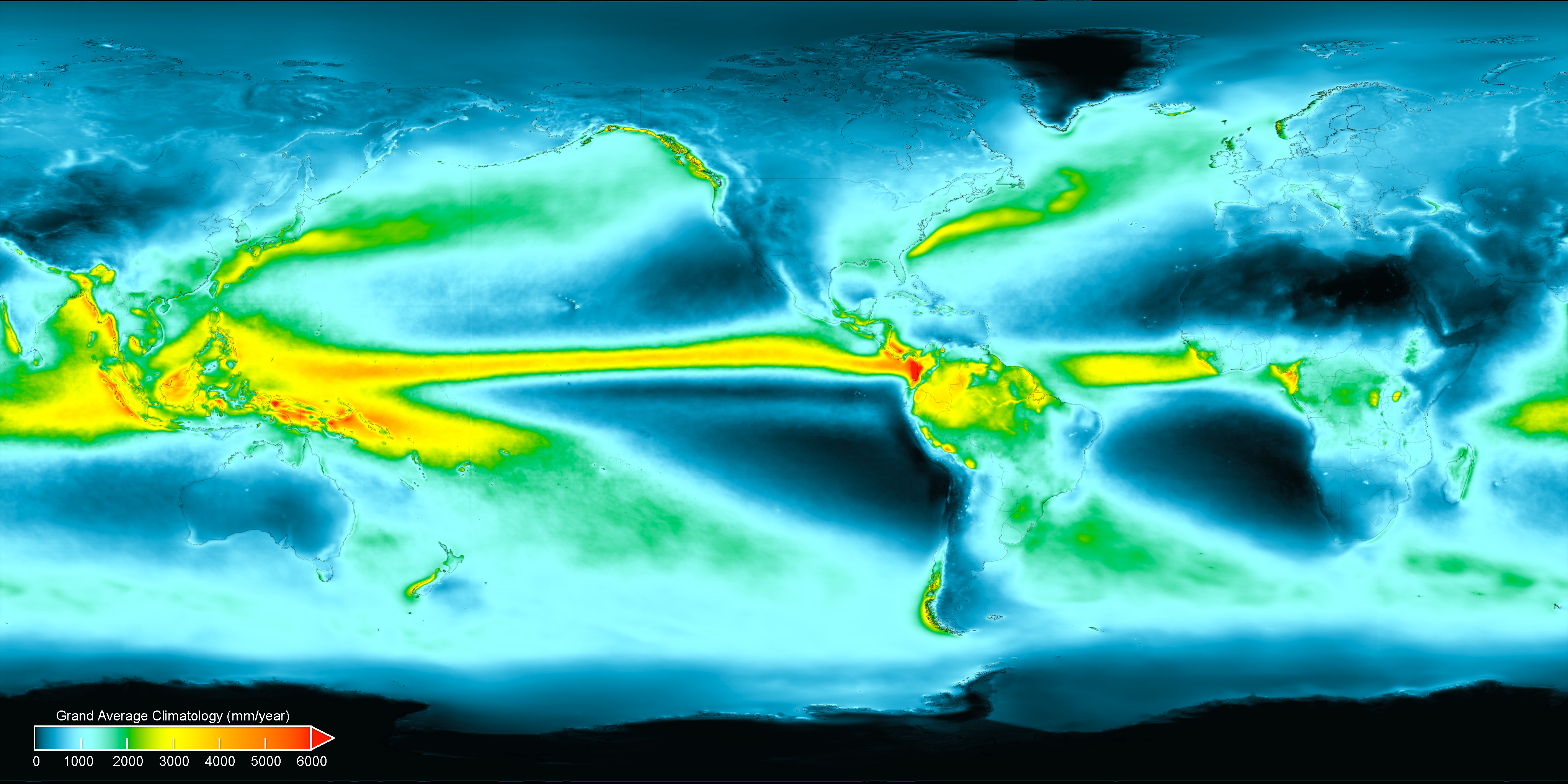}
	\caption{NASA IMERG grand average map computed from June 2000 - May 2023}
	\label{fig:imerggrandavg2024colorbar}
\end{figure}

The map in Figure \ref{fig:my_label1} highlights regions of frequent extreme precipitation under the GNN-$r$P. We see a continuous band of high-risk locations along the equatorial belt and monsoon regions, corresponding to the warm tropics. This aligns with long-term climatology: The NASA IMERG grand average map in Figure \ref{fig:imerggrandavg2024colorbar} shows that annual rainfall is greatest in the Indo-Pacific and Amazon basins, and the frequency is highest along the equator and in monsoon corridors. The equatorial band in Figure \ref{fig:riskzonemodemap} echoes this, indicating that the GNN-$r$P model captures the known spatial focus of tropical extremes. Notably, risk zones also emerge over parts of South and Southeast Asia (monsoon regions) and Central Africa. These are the areas where our model predicts extremes most consistently over the 2000-2021 period, reflecting robust zonal convergence with observational studies of tropical rainfall variability \autocite{allan_advances_2020, knutti_robustness_2013}.

\subsection{Interannual Variability of Spatial Extreme Precipitation Embeddings}
\label{sec:mahalanobis-analysis}

To quantify interannual changes in latent representations, let $\mathbf{z}_j^{(y)}$ and $\mathbf{z}_j^{(y+1)}$ be the embeddings of node $j$ in consecutive years $y$ and $y+1$. 
Define the difference matrix $\mathbf{D} = \mathbf{Z}^{(y)} - \mathbf{Z}^{(y+1)}$. 
The empirical covariance of the differences $\Sigma_\Delta = \mathrm{Cov}(\mathbf{D})$ (with small regularization for numerical stability) is used to compute the Mahalanobis distance:
\[
d_j^{(y,y+1)} = \sqrt{ (\mathbf{z}_j^{(y)} - \mathbf{z}_j^{(y+1)})^\top
	\Sigma_\Delta^{-1} (\mathbf{z}_j^{(y)} - \mathbf{z}_j^{(y+1)}) }.
\]
To evaluate clustering agreement across years we use standard measures such as precision and recall for a chosen high-risk cluster and the adjusted Rand index (ARI) for overall partition similarity \parencite{hubert_comparing_1985}. When interested in tail-index estimates of the scalar scores \(\{r_i\}\), classical estimators such as Hill's estimator may be employed \parencite{hill_simple_1975}.

Mahalanobis distance has been widely applied in spatial ecology and risk studies due to its robustness in multivariate anomaly detection \autocite{conte_spatio-temporal_2016,lotterhos_novel_2021}. In the context of our GNN-$r$P framework, it enables the quantification of year-to-year variability in tail-driven embedding structures.

\begin{figure}[!p]
	\centering
	\includegraphics[angle=90, width=1.6\textheight, height=1.35\textwidth, keepaspectratio]{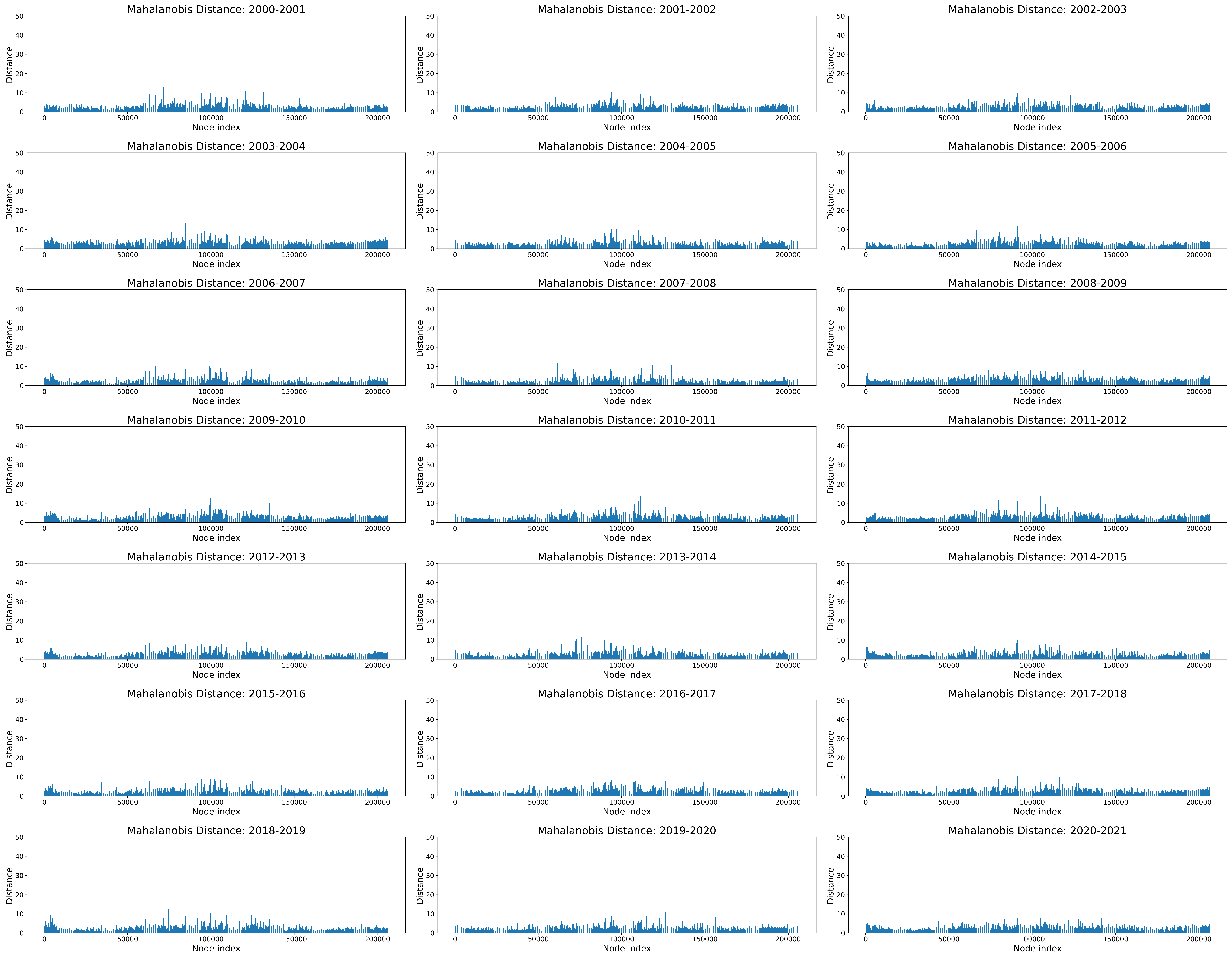}
	\caption{The bar plots of Mahalanobis distance for each pair of consecutive years (2000--2021).}
	\label{fig:mahalanobis_bars}
\end{figure}

Figure~\ref{fig:mahalanobis_bars} presents a mosaic of bar plots showing the distribution of Mahalanobis distance across all spatial nodes in consecutive years from 2000 to 2021. Across most transitions, distances cluster around a mean of approximately 3.0, indicating relative stability in the embedding landscape. However, three pairs stand out with noticeably elevated averages: 2003-2004 (3.55), 2008-2009 (3.58), and 2011-2012 (3.30). These transitions suggest episodes of substantial restructuring in the spatial risk profiles.

These spikes in Figure~\ref{fig:mahalanobis_bars} correspond to well-documented hydrometeorological shifts. For instance, the 2003-2004 transition followed the severe European and Australian droughts of 2003 and coincided with record-breaking monsoon floods across South Asia in 2004 \autocite{dineva_climate-driven_2013, organisation_meteorologique_mondiale_wmo_2005}. Similarly, the 2008-2009 period was marked by successive tropical cyclones and extreme floods across Asia, aligning with the high Mahalanobis distances observed for many nodes. The 2011-2012 transition also shows above-average volatility, consistent with widespread La Niña-related anomalies. These results highlight that the GNN embeddings are sensitive to major reorganizations of the precipitation risk landscape \autocite{zscheischler_future_2018}. By contrast, the 2014-2015 pair shows the lowest mean distance (2.90) and a narrow distribution, suggesting exceptionally stable spatial representations. This coincides with relatively subdued extreme precipitation activity across tropical and monsoonal zones during that time.  

\clearpage

\begin{table}[!h]
	\centering
	\caption{Summary statistics of Mahalanobis distances and corresponding $p$-value distributions across consecutive years.}
	\label{tab:mahalanobis_pvalues} 
	
	\rotatebox{90}{
		
		\begin{tabular}{lrrrrrrrrr|rrrrrr}
			\toprule  
			Year Pair & Mean & Std & Min & 25\% & Median & 75\% & 90\% & 99\% & Max & Mean $p$ & Median $p$ & Min $p$ & Max $p$ & Prop Sig. \\
			\midrule  
			2000--2001 & 3.0486 & 1.5209 & 0.3275 & 2.0850 & 2.7851 & 3.5505 & 4.6281 & 8.8257 & 23.4110 & 0.4746 & 0.4576 & 0.0000 & 1.0000 & 0.1088 \\
			2001--2002 & 3.1406 & 1.4067 & 0.5015 & 2.2020 & 2.9172 & 3.6908 & 4.6138 & 8.4822 & 19.8595 & 0.4348 & 0.3853 & 0.0000 & 1.0000 & 0.1117 \\
			2002--2003 & 3.1482 & 1.4619 & 0.2263 & 2.1784 & 2.9081 & 3.7757 & 4.8484 & 8.3633 & 19.6748 & 0.4366 & 0.3901 & 0.0000 & 1.0000 & 0.1373 \\
			2003--2004 & 3.5489 & 1.3613 & 0.5731 & 2.6772 & 3.3533 & 4.1565 & 5.1375 & 8.0634 & 24.5984 & 0.3010 & 0.1882 & 0.0000 & 1.0000 & 0.1836 \\
			2004--2005 & 3.0815 & 1.4461 & 0.2707 & 2.1352 & 2.8358 & 3.6615 & 4.6203 & 8.4570 & 23.3783 & 0.4553 & 0.4294 & 0.0000 & 1.0000 & 0.1114 \\
			2005--2006 & 2.9905 & 1.5468 & 0.3545 & 1.9581 & 2.7028 & 3.5730 & 4.7011 & 8.7000 & 24.7484 & 0.4971 & 0.5041 & 0.0000 & 1.0000 & 0.1166 \\
			2006--2007 & 2.8602 & 1.5393 & 0.1996 & 1.8626 & 2.6071 & 3.4095 & 4.6023 & 8.4891 & 27.8201 & 0.5301 & 0.5587 & 0.0000 & 1.0000 & 0.1074 \\
			2007--2008 & 2.8313 & 1.4221 & 0.3768 & 1.9430 & 2.4866 & 3.3306 & 4.4307 & 8.1722 & 37.0154 & 0.5477 & 0.6267 & 0.0000 & 1.0000 & 0.0963 \\
			2008--2009 & 3.5781 & 1.4409 & 0.8229 & 2.6375 & 3.3248 & 4.1673 & 5.1504 & 8.7696 & 35.8646 & 0.3045 & 0.1987 & 0.0000 & 1.0000 & 0.1841 \\
			2009--2010 & 3.1967 & 1.5586 & 0.5844 & 2.1096 & 3.0382 & 3.7874 & 4.9164 & 8.8120 & 22.1375 & 0.4263 & 0.3232 & 0.0000 & 1.0000 & 0.1302 \\
			2010--2011 & 3.2038 & 1.4540 & 0.5551 & 2.2054 & 2.9268 & 3.8193 & 4.8643 & 8.4276 & 27.9340 & 0.4264 & 0.3802 & 0.0000 & 1.0000 & 0.1345 \\
			2011--2012 & 3.2993 & 1.5468 & 0.6465 & 2.2338 & 3.0611 & 3.9306 & 4.9774 & 8.8820 & 25.1575 & 0.4025 & 0.3120 & 0.0000 & 1.0000 & 0.1485 \\
			2012--2013 & 3.1329 & 1.5426 & 0.4256 & 2.0554 & 2.9296 & 3.7861 & 4.8715 & 8.4984 & 21.3752 & 0.4459 & 0.3787 & 0.0000 & 1.0000 & 0.1318 \\
			2013--2014 & 3.0707 & 1.5454 & 0.4402 & 1.9893 & 2.8337 & 3.7330 & 4.7755 & 8.5858 & 31.8470 & 0.4682 & 0.4305 & 0.0000 & 1.0000 & 0.1233 \\
			2014--2015 & 2.9020 & 1.4677 & 0.2776 & 1.9420 & 2.5986 & 3.4859 & 4.5215 & 8.2795 & 47.7169 & 0.5202 & 0.5635 & 0.0000 & 1.0000 & 0.1032 \\
			2015--2016 & 2.8796 & 1.5348 & 0.3044 & 1.8617 & 2.5915 & 3.4466 & 4.6439 & 8.3955 & 44.2812 & 0.5297 & 0.5676 & 0.0000 & 1.0000 & 0.1103 \\
			2016--2017 & 3.1279 & 1.4714 & 0.4119 & 2.1407 & 2.8331 & 3.7535 & 4.8355 & 8.2836 & 36.4801 & 0.4506 & 0.4309 & 0.0000 & 1.0000 & 0.1347 \\
			2017--2018 & 3.1882 & 1.3961 & 0.4761 & 2.2406 & 2.9715 & 3.8433 & 4.7798 & 8.1338 & 23.6325 & 0.4172 & 0.3569 & 0.0000 & 1.0000 & 0.1348 \\
			2018--2019 & 2.9605 & 1.4990 & 0.3424 & 1.9438 & 2.6815 & 3.5657 & 4.7181 & 8.2376 & 20.3147 & 0.5022 & 0.5162 & 0.0000 & 1.0000 & 0.1165 \\
			2019--2020 & 3.0122 & 1.3796 & 0.3675 & 2.0736 & 2.7664 & 3.5867 & 4.5801 & 7.9352 & 22.8367 & 0.4762 & 0.4681 & 0.0000 & 1.0000 & 0.1079 \\
			2020--2021 & 3.1098 & 1.4415 & 0.4089 & 2.1265 & 2.8949 & 3.7284 & 4.7388 & 8.1725 & 26.7073 & 0.4453 & 0.3972 & 0.0000 & 1.0000 & 0.1232 \\
			\bottomrule 
		\end{tabular}
	}
\end{table}
\clearpage

Table~\ref{tab:mahalanobis_pvalues} summarizes the Mahalanobis distance statistics together with the distribution of corresponding $p$-values across consecutive years and the proportion of nodes showing statistically significant structural change ($p<0.01$). Several consistent patterns emerge. Mean distances across year pairs lie mostly between about 2.8 and 3.6, indicating persistent but moderate year-to-year adjustments in the embedding landscape rather than wholesale collapses or complete stability. Notably, the high distances observed for some grid cells signal the presence of a small subset of nodes that undergo exceptionally large structural shifts. The minima are not near zero for any year pair, implying that virtually all nodes undergo at least some non-negligible change annually.

The $p$-value summaries provide complementary information about statistical significance: mean and median $p$-values typically lie in the 0.30-0.55 range, indicating that most node-level changes are not individually highly significant under the \(\chi^2\) approximation. Nevertheless, every year pair contains a nontrivial subset of nodes with \(p<0.01\), and the ``Prop Sig." column quantifies this fraction. On average roughly 10-14\% of nodes per year are flagged as statistically significant at \(p<0.01\), while two transitions, 2003-2004 and 2008-2009, stand out with markedly higher shares (approximately 18\%). These particular year pairs combine elevated mean distances, heavy upper tails and larger proportions of significant nodes, consistent with more widespread structural reorganization of the spatial risk landscape in those periods.

\begin{figure}[hbt!]
	\centering 
	\includegraphics[
	width=1.0\textwidth,  
	height=0.7\paperheight, 
	trim=0 0 0 0,
	clip]{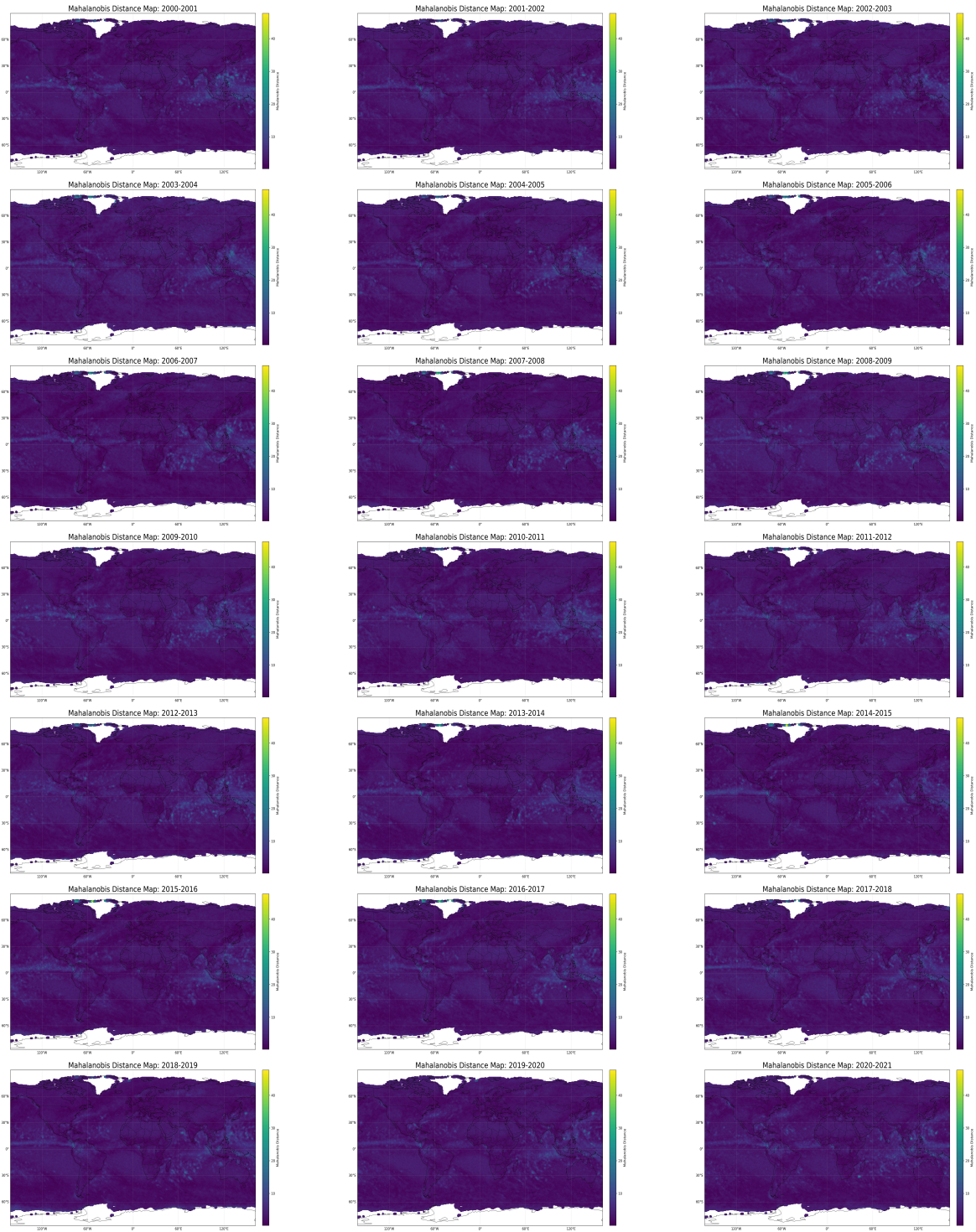}
	\caption{The map of Mahalanobis distance for each pair of consecutive years (2000--2021).}
	\label{fig:mahalanobis_map}
\end{figure}

To further highlight the spatial heterogeneity of these dynamics, Figure~\ref{fig:mahalanobis_map} displays a series of Mahalanobis distance maps. The 2003-2004 panel reveals pronounced hotspots across South Asia, consistent with severe monsoon floods. The 2008-2009 map shows strong anomalies in Southeast Asia and the Caribbean, aligning with cyclone-driven extremes. Elevated distances in western North America during 2011-2012 coincide with La Niña-related drought and flood patterns. These localized signatures provide evidence that the embedding shifts captured by the GNN-$r$P framework reflect real-world structural changes in climate risk regimes.

In combination with the zonation results in Section~\ref{subsec:zonation}, the Mahalanobis analysis shows that core high-risk regions remain persistent, while peripheral areas exhibit higher volatility. This provides a principled mechanism to distinguish stable from transitional phases in the evolution of compound precipitation risk.  

\subsection{Model Comparison}
To evaluate the effectiveness of our proposed GNN-$r$P framework in capturing spatially distributed extreme precipitation risks, this section compares its performance against two benchmark methods. The first baseline, the \textbf{Tail Functional} model, uses GNN-encoded features but defines risk zones directly based on mean tail scores without additional normalization. The second, the \textbf{Top-K Tail} heuristic, selects the top $K$ grid cells with the highest tail scores each year. These baselines represent commonly used alternatives in spatial risk zoning, allowing us to assess the added value of the $r$-Pareto tail normalization and dependence-aware embeddings.

The models are evaluated on annual risk zoning maps from 2000 to 2021 using mean precision, mean recall, and the adjusted Rand index (ARI) as summary metrics. Precision and recall are computed by treating the high-risk zone in a fixed baseline year (e.g., 2015) as the reference label and comparing predicted high-risk labels in other years. The ARI quantifies overall clustering agreement across years regardless of specific label identities, providing a measure of temporal stability in the spatial partitions rather than pointwise prediction accuracy.

\clearpage

\begin{table}[hbt!]
	\centering
	\caption{Performance comparison across three models.}
	\label{tab:model_comparison}
	\begin{tabular}{lccc}
		\toprule
		\textbf{Model} & \textbf{Mean Precision} & \textbf{Mean Recall} & \textbf{Mean ARI} \\
		\midrule
		GNN + Tail Functional & 0.568 & 0.484 & 0.3646 \\
		GNN + $r$-Pareto (ours) & \textbf{0.819} & \textbf{0.542} & 0.3649 \\
		GNN + Top-K Tail & 0.568 & 0.485 & \textbf{0.3656} \\
		\bottomrule
	\end{tabular}
\end{table}

As shown in Table~\ref{tab:model_comparison}, the GNN-$r$P framework substantially improves both precision and recall relative to the two baselines. Specifically, mean precision rises from approximately $0.568$ for the Tail Functional and Top-K Tail models to $0.819$ under the $r$-Pareto model, an absolute gain of about $0.25$. Mean recall increases from $0.484$ and $0.485$ to $0.542$, an absolute improvement of roughly $0.056$--$0.057$. These results indicate that our approach not only more reliably flags the most severe grid cells but also recovers a larger fraction of spatially dispersed high-risk locations that constitute compound extremes.

In contrast, the three models exhibit nearly identical mean ARI values (all around $0.365$), suggesting comparable temporal consistency in clustering partitions. This implies that while $r$-Pareto enhances pointwise identification of high-risk cells (precision and recall), the global latent partition structure across years remains broadly similar across methods. Notably, the Top-K Tail heuristic achieves the numerically largest mean ARI by a small margin.

Overall, these findings highlight the practical value of integrating multivariate tail normalization and dependence-aware spatial embeddings into risk zoning. The GNN-$r$P framework effectively combines nonlinear spatial representations with principled tail parametrization to improve both precision and coverage in high-risk area detection, which is crucial for applications in climate risk assessment and adaptation planning \autocite{koh_using_2024,de_fondeville_high-dimensional_2018}.

\subsection{Temporal Consistency of Risk Zoning under SSP5-8.5}
To investigate how spatial precipitation risk patterns may evolve under a high-emission climate scenario, we apply the GNN-$r$P framework to the SSP5-8.5 projection dataset. The SSP5-8.5 scenario represents a high greenhouse gas concentration pathway commonly used for climate impact assessments, providing a basis to explore potential future changes in extreme precipitation. By analyzing annual risk zoning from 2030 to 2074, we aim to quantify both the persistence of established high-risk regions and the emergence of new hazard hotspots under a warming climate.

\begin{figure}[hbt!]
	\centering
	\includegraphics[width=0.8\linewidth]{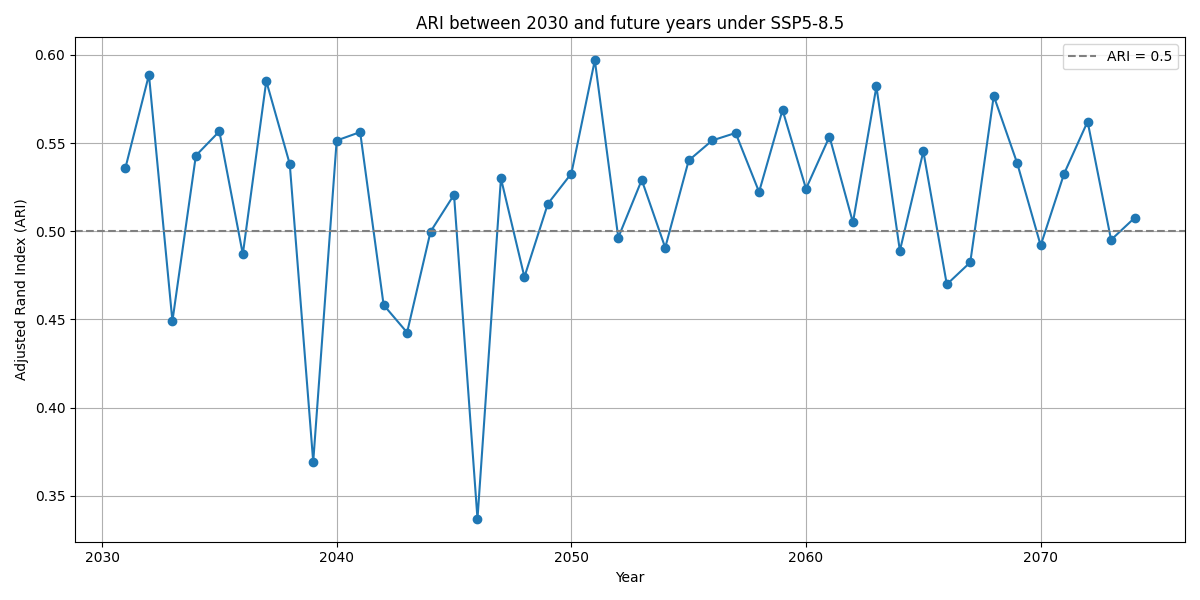}
	\caption{Adjusted Rand Index (ARI) between 2030 and subsequent years under SSP5-8.5. The ARI measures the similarity between annual spatial risk clusterings.}
	\label{fig:ssp58ari}
\end{figure}

We first compute the Adjusted Rand Index (ARI) between each year's clustering and the 2030 baseline, as shown in Figure~\ref{fig:ssp58ari}. The ARI values quantify how similar the spatial partition of extreme precipitation risk is from year to year, providing a compact measure of interannual persistence and reorganization. Overall, the ARI series exhibits moderate to high similarity with nontrivial fluctuations. Early in the projection period, the ARI remains relatively high: the ARI between 2030 and 2031 equals about $0.536$, rising to $0.589$ for 2032. Several additional local maxima occur later, most notably the largest observed value $0.597$ at 2051, with secondary peaks around 2037 ($0.585$), 2063 ($0.582$), and 2068 ($0.577$). These elevated values indicate years when the clustering arrangement remains strongly similar to the 2030 baseline, reflecting persistence of core risk zones across multi-year intervals.

The ARI trajectory is not monotonic, however. Distinct troughs appear, with the lowest ARI of $0.337$ for the 2030 to 2046 comparison, and other pronounced dips for 2030 to 2039 ($0.369$) and 2030 to 2033 ($0.449$). Such troughs denote intervals of substantial reorganization in the zonation, potentially driven by major hydrometeorological anomalies, shifts in large-scale climate modes, or nonlinear interactions that alter regional precipitation extremes. The coexistence of peaks and troughs suggests that while the spatial risk structure is broadly persistent on decadal scales, episodic reconfigurations punctuate this stability.

\begin{figure}[hbt!]
	\centering
	\includegraphics[width=0.8\linewidth]{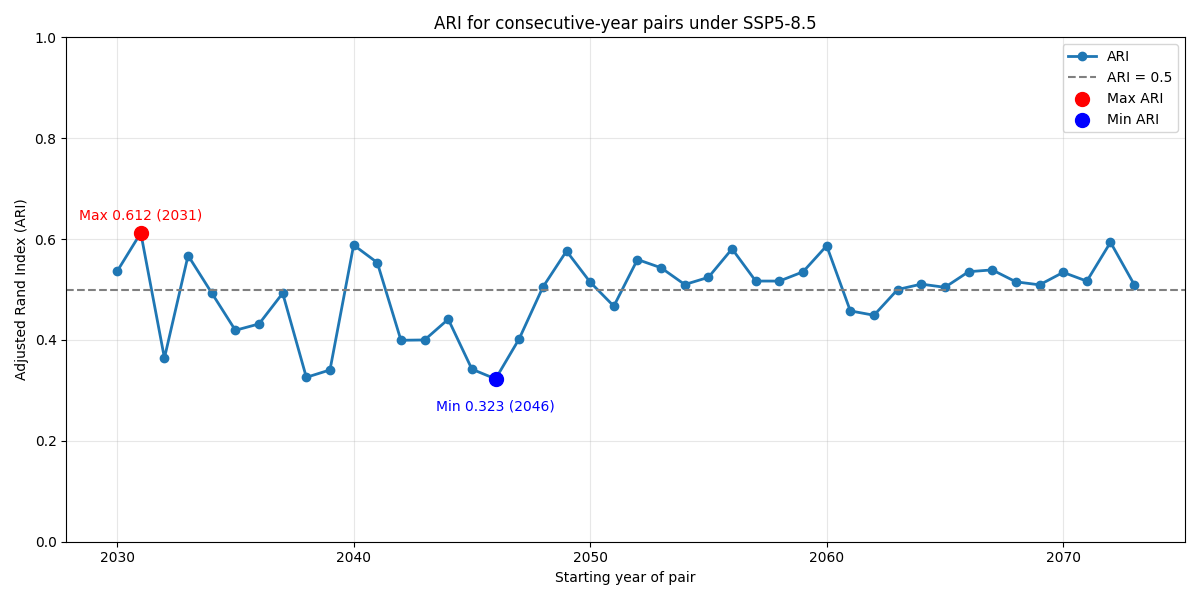}
	\caption{Adjusted Rand Index (ARI) for consecutive-year pairs (2030-2031 to 2073-2074) under SSP5-8.5. The dashed gray line denotes ARI = 0.5 (moderate similarity). Red and blue markers indicate the maximum (0.612, 2031-2032) and minimum (0.323, 2046-2047) ARI values, respectively.}
	\label{fig:consecutive_ari}
\end{figure}

To further capture interannual dynamics, Figure~\ref{fig:consecutive_ari} presents ARI values for consecutive-year pairs (e.g., 2030-2031, 2031-2032). The visualization employs a line plot with circular markers to trace fluctuations across the 2030-2074 horizon, supplemented by reference markers to emphasize key extremes. A dashed gray line at ARI = 0.5 serves as a benchmark for moderate similarity. Red and blue markers identify the global maximum and minimum, with annotations showing their magnitudes and corresponding years. Specifically, the highest ARI ($0.612$) occurs for the 2031-2032 pair, reflecting exceptional continuity in spatial risk zoning, while the lowest ARI ($0.323$) is observed for 2046-2047, indicating substantial reorganization. These features illustrate that while spatial risk patterns remain moderately stable overall, notable disruptions punctuate the temporal sequence.

\begin{figure}[hbt!]
	\centering
	\includegraphics[width=0.75\linewidth]{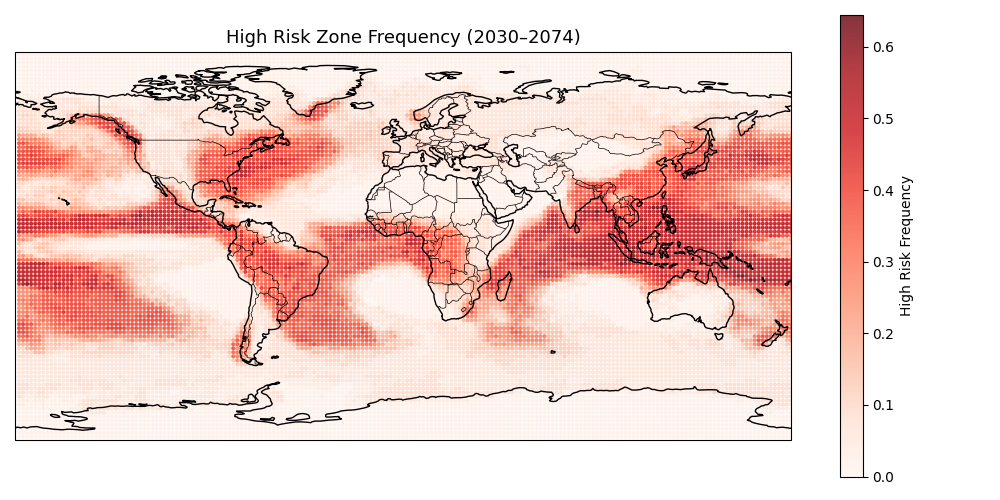}
	\caption{Frequency of classification into Zone 3 (highest risk) under SSP5-8.5 from 2030 to 2074. Darker red indicates higher temporal persistence of extreme precipitation risk.}
	\label{fig:ssp58freqmap}
\end{figure}

To complement the ARI-based assessment of temporal consistency, we examine cumulative persistence of high-risk zones. Figure~\ref{fig:ssp58freqmap} visualizes how frequently each grid cell is classified into Zone 3 (highest risk) across 2030-2074, with darker red indicating more persistent exposure to extreme precipitation. The map highlights several high-persistence regions: the Indo-Pacific Warm Pool, equatorial West and Central Africa, and parts of the Amazon basin, all showing Zone 3 frequencies exceeding 0.5. These areas are climatically prone to convective intensification and are projected to experience repeated extreme rainfall events under warming scenarios \autocite{RN24}.

In contrast, mid-latitude continental interiors, high-latitude regions, and subtropical dry zones show low persistence, consistent with lower baseline precipitation variability. Emerging hotspots over the equatorial oceans and maritime Southeast Asia may reflect changing ENSO\footnote{ENSO: El Niño-Southern Oscillation} regimes or shifts in zonal circulation. Together, the ARI and frequency-based analyses demonstrate that the GNN-$r$P framework captures both persistent and evolving spatial structures of compound precipitation risk under the SSP5-8.5 scenario, providing a useful tool for climate risk assessment and adaptation planning \autocite{hubert_comparing_1985,steinley_properties_2004}.

\section{Conclusion}\label{sec:conclusion}
This study proposes a hybrid framework combining graph neural networks and $r$-Pareto processes to model spatial patterns of extreme precipitation under both historical and future climate scenarios. By constructing spatial graphs and learning nonlinear low-dimensional embeddings of local precipitation features, this approach captures nonstationary dependence structures and identifies coherent high-risk zones based on robust tail detection. Compared with frequency-based or heuristic top-K methods, the GNN-$r$P model achieves superior clustering accuracy (mean precision = 0.819), improved spatial coherence, and resilience to data noise, making it well-suited for real-world climate analytics.

Empirical results highlight consistent high-risk zones across tropical and subtropical regions, particularly in the Caribbean, Southeast Asia, equatorial Africa, and the Pacific Islands, where climate extremes are both frequent and consequential. Scenario analysis under SSP5-8.5 suggests intensifying and increasingly uncertain risk patterns beyond 2060, emphasizing the urgency of adaptive and structure-aware modeling approaches.

Importantly, our findings carry practical implications for regional adaptation and risk financing. At-risk nations such as Haiti, Mozambique, and Bangladesh, which face both climatic exposure and institutional constraints, would benefit from spatially resolved modeling for targeting investments in resilient infrastructure, early-warning systems, and adaptive insurance mechanisms. Regional parametric insurance schemes such as CCRIF\footnote{CCRIF: Caribbean Catastrophe Risk Insurance Facility}, ARC\footnote{ARC: African Risk Capacity Limited}, and PCRAFI\footnote{PCRAFI: Pacific Catastrophe Risk Assessment and Financing Initiative} can be improved using machine learning informed tail risk metrics to support more equitable and timely payouts. Furthermore, global climate funds such as the UN Loss and Damage Fund or the Global Shield Against Climate Risks should leverage fine-grained risk models to allocate resources in a more risk-sensitive manner, beyond crude GDP thresholds.

Despite its strengths, several limitations warrant attention. First, the model's output can be sensitive to the choice of tail thresholds and embedding dimensionality, potentially affecting interpretability near climatic transition zones. Second, while the framework captures localized spatial dependencies, it does not explicitly model long-range teleconnections (e.g., ENSO), which may drive interannual variability. Third, the model currently assumes static spatial graphs and independent years, overlooking temporal autocorrelation or compound extremes involving other hazards such as heatwaves or cyclones. Future work could extend the model with dynamic graph structures, climate indices, and multi-hazard coupling to better reflect the complex nature of climate risks.

In conclusion, our GNN + $r$-Pareto approach provides a promising pathway for data-driven risk zoning and climate adaptation planning, blending machine learning and extreme value theory in a scalable, interpretable framework. The integration of it opens new possibilities for compound risk analysis, regional adaptation planning, climate model validation and catastrophe insurance design. By linking spatial analytics with actionable risk insights, this method lays the technical foundation for globally coordinated adaptation strategies aligned with the Sendai Framework and the Paris Agreement.

\printbibliography

\end{document}